\documentclass[twocolumn,amsmath, amssymb, superscriptaddress, pra]{revtex4}
\usepackage{graphicx}
\usepackage{dcolumn}
\usepackage{bm}
\usepackage{epsf}
\usepackage{amssymb}
\usepackage{subfigure}
\usepackage{cancel}
\DeclareGraphicsExtensions{.eps}
\begin{document}


\author{David S. Simon}
\email[e-mail: ]{simond@bu.edu} \affiliation{Dept. of Physics and Astronomy, Stonehill College, 320 Washington Street, Easton, MA 02357} \affiliation{Dept. of
Electrical and Computer Engineering \& Photonics Center, Boston University, 8 Saint Mary's St., Boston, MA 02215, USA}
\author{Casey A. Fitzpatrick}
\email[e-mail: ]{cfitz@bu.edu} \affiliation{Dept. of Electrical and Computer Engineering \& Photonics Center, Boston University, 8 Saint Mary's St., Boston, MA
02215, USA}
\author{Shuto Osawa}
\email[e-mail: ]{sosawa@bu.edu} \affiliation{Dept. of Electrical and Computer Engineering \& Photonics Center, Boston University, 8 Saint Mary's St., Boston, MA
02215, USA}
\author{Alexander V. Sergienko}
\email[e-mail: ]{alexserg@bu.edu} \affiliation{Dept. of Electrical and Computer Engineering \& Photonics Center, Boston University, 8 Saint Mary's St., Boston,
MA 02215, USA} \affiliation{Dept. of Physics, Boston University, 590 Commonwealth Ave., Boston, MA 02215, USA}

\begin{abstract}

It is shown that quantum walks on one-dimensional arrays of special linear
optical units allow the simulation of discrete-time Hamiltonian systems
with distinct topological phases. In particular, a slightly modified
version of the Su-Schrieffer-Heeger (SSH) system can be simulated, which
exhibits states of nonzero winding number and has topologically-protected
boundary states. In the large-system limit this  approach uses
quadratically fewer resources to carry out quantum simulations than
previous linear-optical approaches and can be readily generalized to
higher-dimensional systems. The basic optical units that implement this
simulation consist of combinations of novel optical multiports that allow
photons to reverse direction.
\end{abstract}


\title{Quantum Simulation of Topologically Protected States Using Directionally-Unbiased Linear Optical Multiports}
\maketitle

\section{Introduction}

The rapidly expanding research activity currently underway on quantum computing \cite{nielsen,kitaev,jaeger,falci}  is ultimately an outgrowth of Richard
Feynman's observation that quantum systems are necessary to efficiently simulate other quantum systems \cite{feyn,aspuru,johnson}. The goal of quantum simulation
is therefore to find simple quantum systems that can accurately and efficiently simulate specific properties of interest in more complex quantum physical
entities.

The behavior of a quantum system arises from interference between multiple
solutions of a linear wave equation. This can be seen most clearly in
Feynman's path integral formalism \cite{path1,path2}, where the observable
output state is a linear superposition of all allowed intermediate
trajectories. In a similar manner, linear-optical systems make use of
interference between light waves that arise as solutions to the linear
Helmholtz equation. For systems in which particle number is conserved
(electrons in a solid, for example), linear optics would therefore seem to be
a natural resource to exploit in order to carry out quantum simulations. In
particular, photonic quantum walks \cite{kempe,ven} can produce the complex
interference patterns needed for such simulations. Because of the relatively
feeble interactions that photons have with their surroundings, many of the
complexities associated with other physical implementations of quantum
simulations are greatly reduced in an optical setting. In addition, light is
not only easy to produce and detect, but it can be tailor-made with a high
degree of control over its frequency, polarization, and spatial and temporal
profiles. Further, quantum effects are readily visible in optics; for
example, photon pairs can be routinely produced with high degrees of
entanglement \cite{shih,ou,simbook}.

Systems with nontrivial topological behavior arise naturally in the study of solids, as well as in other areas of physics. They lead to wavefunctions with
nonzero Chern number or winding number, to topologically protected edge or boundary states, and to phase transitions between distinct topological states (see
\cite{hasan,kitagawa,asboth,bern,stanescu} for reviews). Alongside theoretical work and experimental implementation, quantum simulation of these behaviors has
also become an active area of current research. For example, simulations have been carried out with ultracold atoms, both in free space and confined to optical
lattices \cite{roust1,roust2,liu,li,jav,zhang1,zhang2}, as well as in photonic quantum walks \cite{broome,kitagawa,kit1,kit2,kit3}.

The approaches used up to now for quantum simulations of topologically-nontrivial physical systems have substantial limitations. For example, working with atoms
requires extremely low temperatures in order to avoid decoherence. This adds numerous complications to the experiments and makes this approach unlikely to be
useful outside of research labs. On the other hand, analogous simulations done with optical quantum walks have their own complications. In particular, they
require a set of optical resources (beam splitters, mirrors, etc.) that grows rapidly with the number of steps in the walk.

These factors strongly limit the ability to use the current optical approaches for practical simulations on a large scale, and so it is of interest to
investigate novel schemes that may be more easily scalable. Here we present a linear-optical strategy whose resource requirements grow at a quadratically slower
rate than previous optical approaches. It is currently practical to carry out a table-top version of this procedure, and in the near future it should be
plausible to implement it on much larger scales by integrating all of the required optical elements onto optical chips that can be fabricated in large numbers
and arranged into the desired configurations with high stability. In contrast to the quadratic growth in previous optical implementations, the resources required
here  scale only linearly with number of steps.   Furthermore, this scheme has the advantage that the parameters of the underlying system on which the walk
occurs can be readily varied to produce a variety of simulated behaviors.

In \cite{simham} a linear-optical method was proposed for using photonic quantum walks to carry out quantum simulations of topologically-trivial nearest-neighbor
Hamiltonians in the context of one-dimensional discrete-time physical models. This was accomplished by means of chains of simple linear optical units. Different
Hamiltonians could be simulated by varying the arrangement of these units, or by varying their internal parameters. By going from a one-dimensional chain to two-
or three-dimensional arrays,  Hamiltonians exhibiting more complicated band gap structures can also be implemented.

In the current paper, the simple periodic lattice of \cite{simham} is replaced by a pair of two interlaced sublattices with different parameters, leading to a
substantial generalization in the types of behaviors attainable. In particular, simulation of topologically non-trivial Hamiltonians become possible, with
features such as nonzero winding number and topologically-protected boundary states.

The basic optical units utilized in this scheme are the directionally-unbiased optical multiports proposed in \cite{threeport}. These devices can be thought of
as scattering centers of the type that have been discussed in the abstract context of optical graph systems \cite{hbf,fh1,fh2,fh3}. In a graph model, an incident
photon is constrained at each time step to scatter into one of a finite number of modes. One of these modes is the time-reversed version of the input mode, so it
is necessary that the multiport allows the photon to reverse direction and exit back out the input port. Such reversible multiports can be constructed using only
linear optics and can be thought of as artificially-created optical ``meta-atoms'', with lattices of them forming a type of metamaterial. In this sense, the
current paper is complementary to work that seeks to produce topological behavior in dielectric metamaterials \cite{slob}.

\begin{figure}
\begin{center}
\subfigure[]{
\includegraphics[scale=.25]{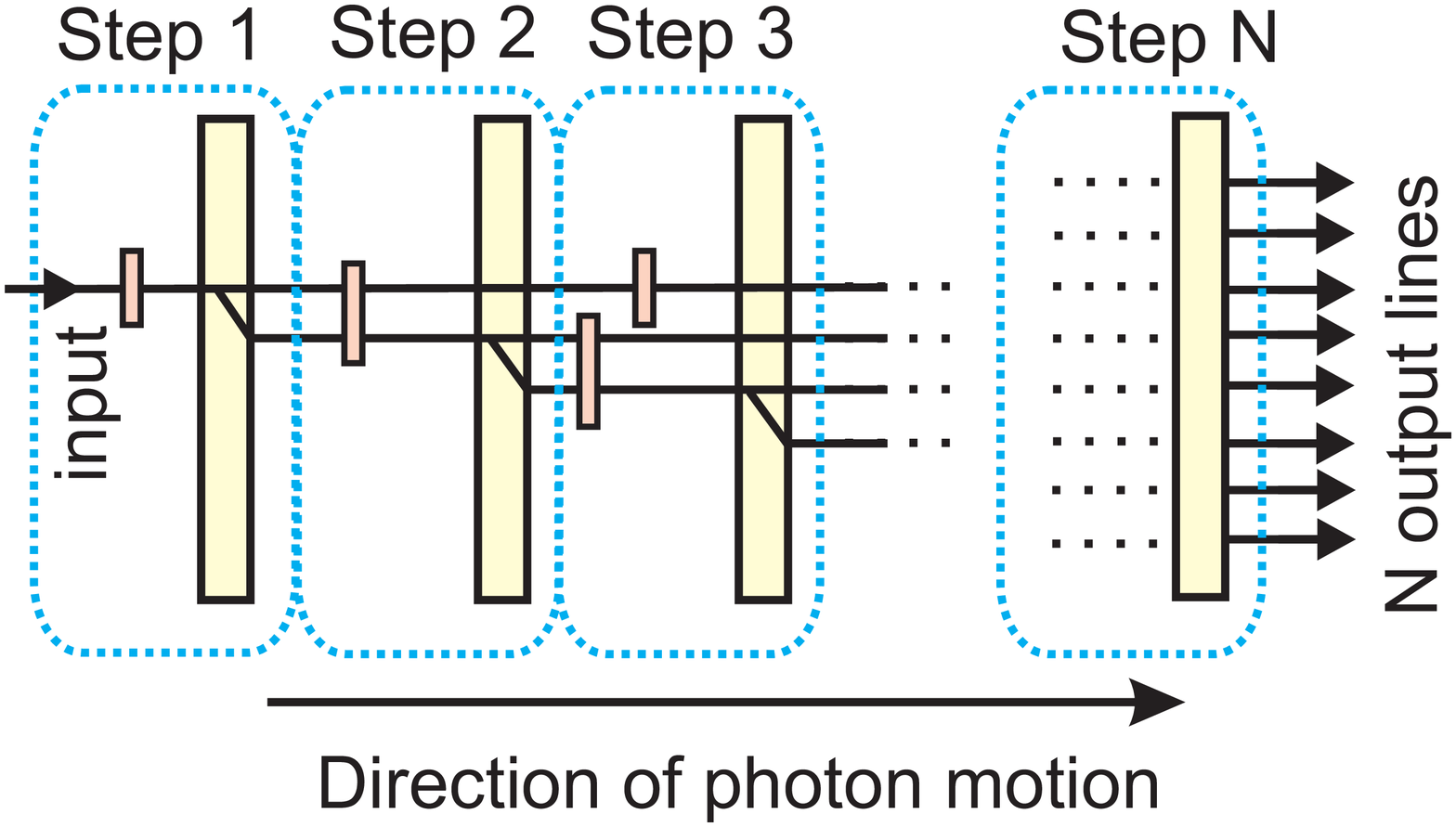}}
\qquad  \subfigure[]{
\includegraphics[scale=.35]{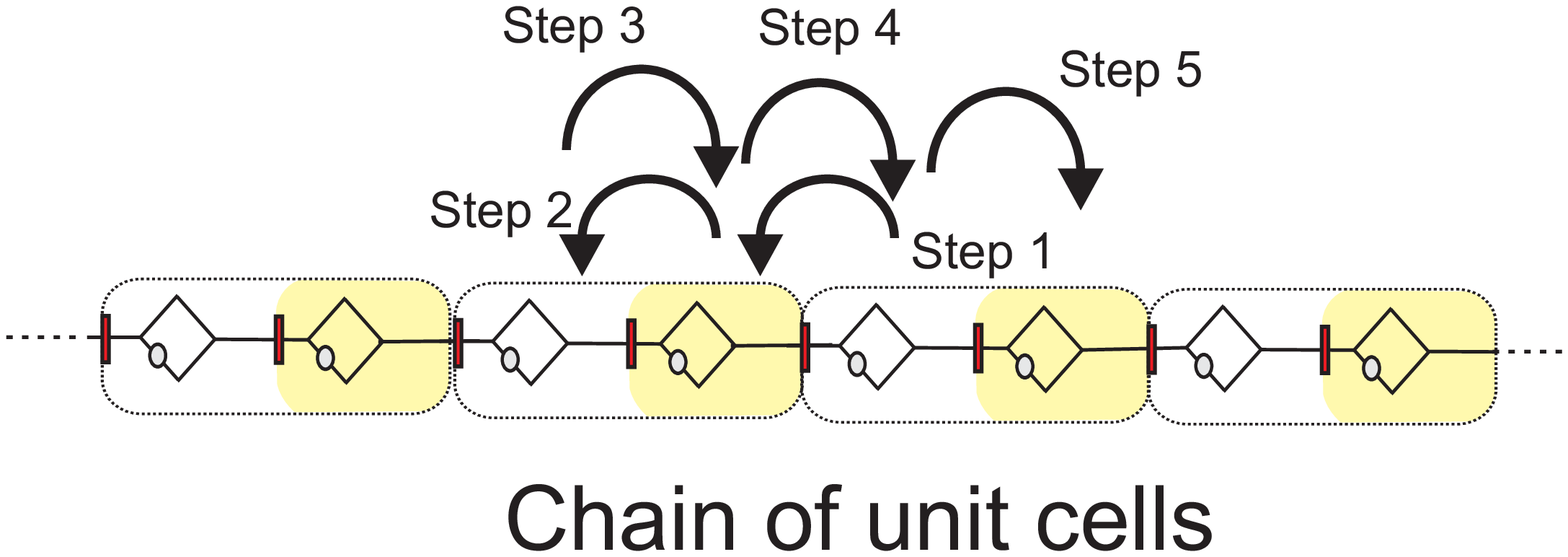}}
\caption{(a) In prior approaches to topological system simulation with linear
optics, the number of optical lines increases with each step and the
walk occurs in the transverse direction, requiring
quadratic increase of resources as the number of steps increases.
(b) The approach using directionally-unbiased multi-ports only requires motion
along a single line to produce the same effect. The quantum walk is in the
longitudinal, rather than the transverse direction, and so only requires linear
resource growth.}\label{explodefig}
\end{center}
\end{figure}

The significant reduction in resources in the current proposal compared to previous optical approaches is a direct consequence of the fact that the input ports
of the unbiased multiport  serve also as output ports. As a result, the flow of photons can reverse direction and traverse the same unit multiple times instead
of needing additional units at each time step. This is illustrated in Figure \ref{explodefig}: previous linear-optical implementations involve a splitting of
optical paths at each step, causing the number of outputs and the number of beam splitters, phase plates, etc. to increase with each step. Although the overall
flow of photons in time is toward right, the quantum walk is occurring in the transverse direction. If $N$ is the number of time steps, then the total resources
grow proportional to $N^2$. Effectively, the standard approach requires a two-dimensional network to carry out a one-dimensional walk. However, with reversible
units of the type used in the current paper, the walk occurs in the longitudinal (horizontal) direction, requiring only a single line with length of order $N$ to
carry out a walk of $N$ steps.  The currently-proposed method also scales up to systems with more spatial or internal degrees of freedom in a straightforward
manner.

We briefly review directionally-unbiased multiports and topologically non-trivial discrete-time Hamiltonian systems in Sections \ref{unbiased} and
\ref{topsection}, respectively, before using the multiports to demonstrate linear optical simulation of topologically protected states in section
\ref{sshsection}. We briefly discuss these results in section \ref{conclude}.

\section{Directionally-Unbiased Multiports}\label{unbiased}

\begin{figure}
\begin{center}
\subfigure[]{
\includegraphics[scale=.2]{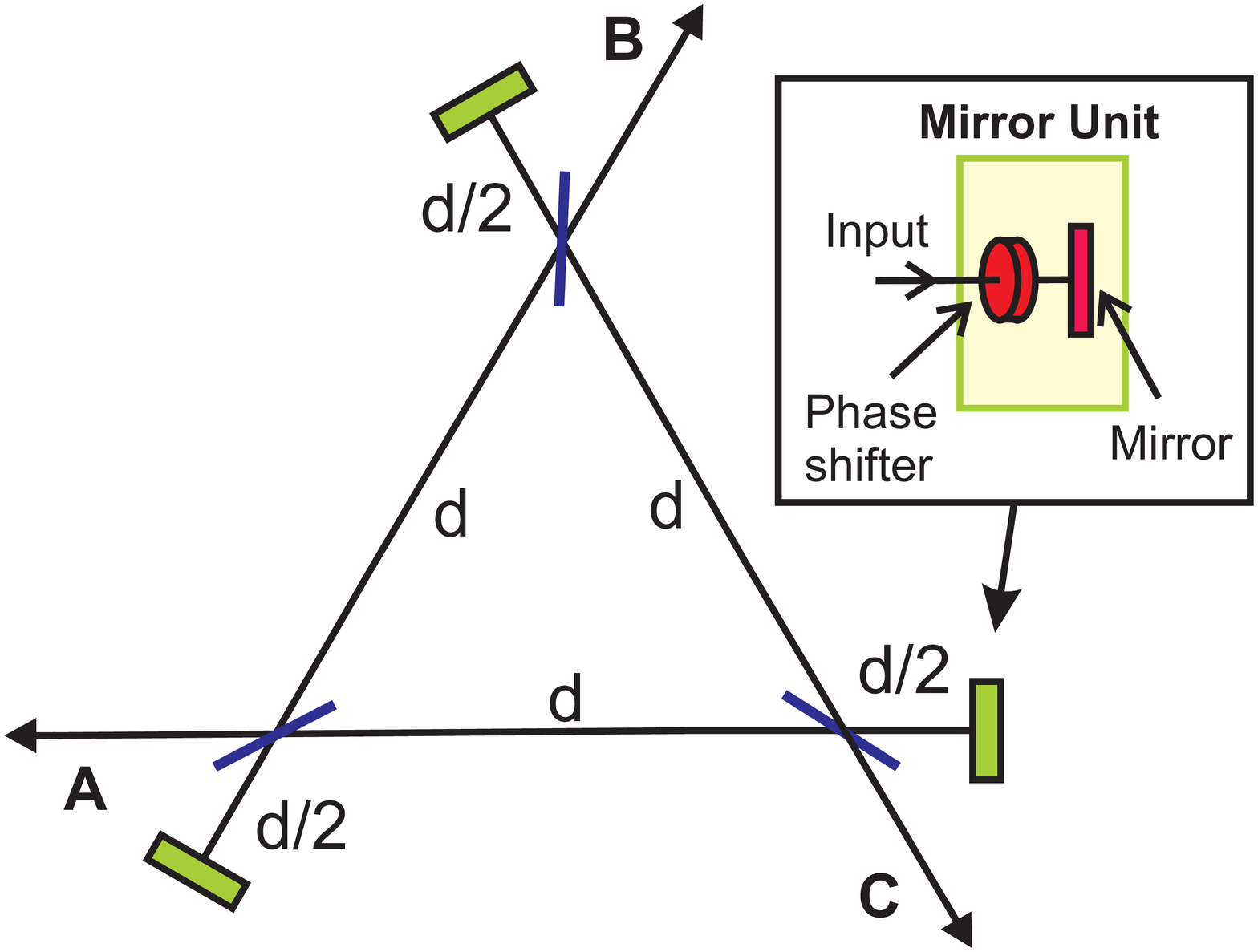}}
\qquad  \subfigure[]{
\includegraphics[scale=.2]{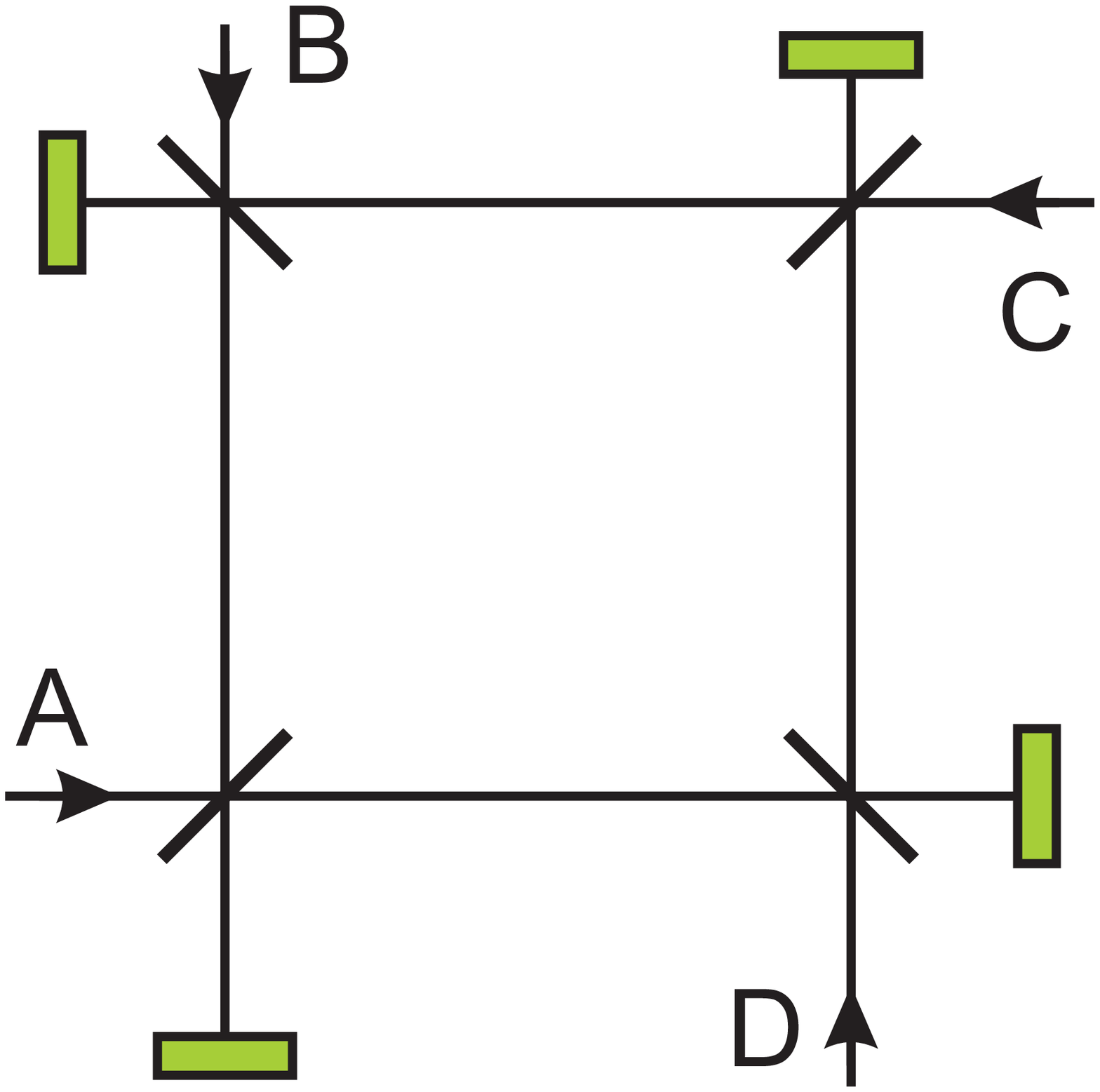}}
\caption{(a) The directionally-unbiased three-port. (b) The directionally-unbiased four-port. The rectangles after the beam splitters in (a) and (b) represent the vertex mirror unit shown in the inset of (a). This unit consists of a mirror and a phase-shifter. The distance between each beam splitter and the adjacent mirror unit is half the distance $d$ between one beam splitter and the next. }\label{nportfig}
\end{center}
\end{figure}

Ordinary beam splitters and their multiport generalizations only allow
one-way movement of photons; the light never reverses direction inside. In
\cite{threeport}, a generalized multiport was proposed which allows such a
reversal. Such a device, called a directionally unbiased multiport, allows
the experimental implementation of scattering-based quantum walks on graphs
\cite{fh1,fh2,fh3}. Examples of unbiased $n$-ports for $n=3$ and $n=4$ are
shown in Fig. \ref{nportfig}(a) and (b). Only the three-port version will be
used in the following.

The directionally unbiased multiports are linear optical devices with the
input/output ports attached to vertex units of the form shown in the inset of
Fig. \ref{nportfig}(a). Each such unit contains a  beam splitter, mirror and
phase shifter.  The beam splitter-to-mirror distance ${d\over 2}$ is half of
the distance $d$ between the vertex units in the multiport. The phase shifter
provides control of the properties of the multiport, since different choices
of phase shift at the vertices affect how the different photon paths through
the device interfere with each other.

If the unit is sufficiently small (quantitative estimates of the required
size and other parameter values may be found in \cite{threeport}) then its
action can be described by an $n\times n$ unitary transition matrix $\hat U$
whose rows and columns correspond to the input and output states at the
ports. If the internal phase shifts at all three mirror units are equal, then
an explicit form of the unitary transition matrix $\hat U$ can be found:
\begin{equation}\hat U={{e^{i\theta}}\over {2+ie^{i\theta}}}\left( \begin{array}{ccc} 1 & ie^{-i\theta}-1 & ie^{-i\theta}-1 \\
ie^{-i\theta}-1 & 1 & ie^{-i\theta}-1\\ ie^{-i\theta}-1 & ie^{-i\theta}-1& 1
\end{array}\right) ,\label{general}\end{equation} where $\theta$ is the total phase shift at each mirror unit (including both the mirror and the phase plate).
The rows and columns refer to the three ports $A$, $B$, $C$.

Two special cases of this result can be noted. First, if the internal phase
shifts at the vertices are set to $\theta ={\pi \over 6}$ then
the exit probabilities at all three ports are equal, but at the cost of having
different phase factors for different transitions. The case where are all the exit probabilities are equal is referred to as the \emph{strictly unbiased} case
\cite{simham}.

A second notable special case of Eq. \ref{general} is  when $\theta
=-{\pi\over 2}$. This choice ensures that all of the photon paths entering
and exiting at any pair of ports will be in phase with each other
\cite{threeport}. The transition amplitude is then always pure imaginary for
every pair of input and output ports, which provides simplifications when
adding multiple transition amplitudes. For this case, which will be the main
focus here, the three-port takes an input state $|\psi_0\rangle$ to an output
state $|\psi\rangle =\hat U|\psi_0\rangle$, where
\begin{equation}\hat U=-{i\over 3}\left(
\begin{array}{ccc} 1 & -2 & -2\\ -2 & 1 & -2\\ -2 & -2& 1
\end{array}\right) .\label{U3port}\end{equation}
The simulation system of section \ref{sshsection} will be built from units
described by Eq. \ref{U3port}.

\section{Winding Numbers and Topologically Protected States}\label{topsection}

\subsection{Topological phases}
The object of study here is a discrete time system, described by a Hamiltonian $\hat H$ and a discrete-time evolution matrix $\hat U =e^{-i\hat HT}$ that takes
the system forward one time-step $T$. (Here, the units are chosen such that $\hbar=1$.) For initial state $|\psi(0)\rangle$, the state at time $t=nT$ is
\begin{equation}|\psi(nT)\rangle =\hat U^n |\psi(0)\rangle .\end{equation} Wedging the matrix $\hat U$ between a pair of desired initial and final states gives the
transition amplitude per time step between those states. We define quasi-momentum $k$ on a one-dimensional periodic lattice made from a sequence of repeating
unit cells. These cells are labeled by an integer, $m$. Since the position variable $m$ is dimensionless and discrete, the quasi-momentum $k$ will be as well. A
single Brillouin zone runs from $0$ to $2\pi$, and $k$ is only conserved modulo $2\pi$.

The Hamiltonian generates time evolution in some space that may include both
spatial and internal degrees of freedom.   As the momentum is varied over the
width of a full Brillouin zone, the parameters defining $\hat H$ will trace
out a closed path in the parameter space. Topological obstructions may
prevent some of these paths from being continuously deformed into each other
as the system parameters vary, leading to distinct topological phases of the
system. In this case, all quantities that are constant on each equivalence
class will be topologically protected and stable under small perturbations.
The distinct phases are usually distinguished by integer-valued quantities
such as the the winding number $\nu$ in one-dimensional lattice systems or
the Chern number two-dimensions.

When two one-dimensional systems with different topological phases are
brought into contact, solutions can only propagate from one region to the
other if they change winding number, which in turn only occurs if the band
gap between quasi-energy levels vanish at the boundary. The closing of the
gap therefore implies the existence of states that are exponentially
localized in the vicinity of the boundary \cite{kitagawa,asboth}, and
continuous variations of the system parameters in the two bulk regions leave
them intact.  These boundary states have been widely studied in recent years
\cite{obuse,asboth2,asob,edge}.

%
%

\subsection{The SSH model.} An example of a Hamiltonian with topological states is the Su-Schreiffer-Heeger (SSH) Hamiltonian \cite{su}, which is used, for
instance, to model  the hopping of electrons along the length of a polyacetylene chain, and which is closely related to a structure appearing in quantum field
theory models \cite{jackiw}.

The SSH system is shown schematically in Fig. \ref{sshfig}. There is a set of lattice sites or cells (labeled by integer $m$), each of which contains two
subsites, denoted as $a$ and $b$ in the figure; these two lattice subsites represent possible ``internal'' states at cell $m$.  There is some amplitude per unit
time $v$ to switch between the two states within the same cell, and an amplitude per time $w$ to hop to the adjacent lattice sites. When the site changes, the
state also flips, and the amplitudes have to be symmetric in the sense that they are the same (up to complex conjugation) for hops to the left and to the right.

The Hamiltonian is of the form:
\begin{eqnarray}\hat H &=& v \sum_{m=1}^N \left( |m,b\rangle \langle m,a| +  |m,a\rangle \langle m,b|  \right)\label{sshhamiltonian} \\
& & + w\sum_{m=1}^{N-1} \left( |m+1,a\rangle \langle m,b| +  |m,b\rangle \langle m+1,a|  \right) ,\nonumber\end{eqnarray} where $N$ is the number of cells in the
chain. $|m,a\rangle$, for example, denotes the state with a particle at site $m$ in substate $a$.

At each fixed lattice site $m$ or each fixed $k$, this Hamiltonian is therefore a two-dimensional matrix, and can be written in terms of the identity matrix and
the Pauli matrices; for example, in momentum space one may write
\begin{equation}\hat H(k) =
d_0(k) I +\bm d(k)\cdot \bm\sigma .\label{Handd}\end{equation}
This describes dynamics in a two dimensional ``internal'' subspace labeled by the two substates present at each lattice site. Generically, the two energy levels
are separated by a $k$-dependent gap.

The insulator described by this Hamiltonian becomes a conductor when the
vector $\bm d(k)$ vanishes; at these points the discrete energy levels meet
and the energy gap between bands vanishes.  In the SSH model $d_0=d_z=0$, so
that the space of possible $\bm d$ values collapses to the two-dimensional
$(d_x,d_y)$ plane. This means that paths encircling the origin cannot be
contracted continuously to a point and have nonzero winding number.  The
winding number is highly stable in the sense that perturbations causing
continuous variations of the parameters cannot stimulate transitions between
topological classes. The SSH system therefore can support highly localized
boundary states at the interfaces between regions of different topological
phase. The appearance of such localized states will be used in the next
section as a signal to verify the existence of distinct topological phases.


\section{Simulating the Modified SSH Hamiltonian Optically}\label{sshsection}

\begin{figure}
\centering
\includegraphics[totalheight=1.1in]{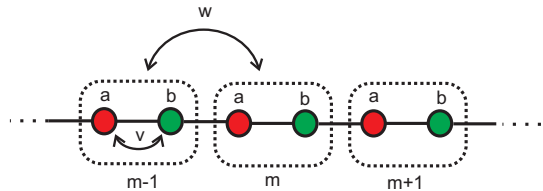}
\caption{The SSH Hamiltonian describes motion of a particle hopping on a chain of sites with two substates per site. $v$ and $w$ are respectively the intracell and intercell hopping amplitudes per unit time.}
\label{sshfig}
\end{figure}

\begin{figure}
\centering
\includegraphics[totalheight=1.2in]{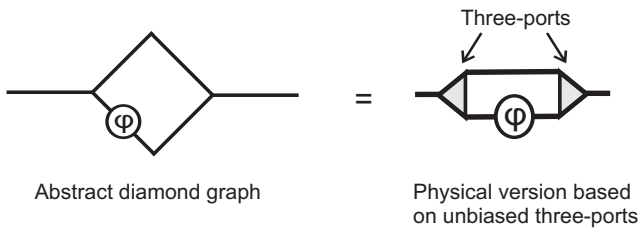}
\caption{The abstract diamond graph \cite{fh1,fh2,fh3} consists of
two three-port scattering vertices connected on two edges,
with a phase shift between them (left). Since the scattering vertices can be
implemented physically by unbiased three-ports, the diamond graph is
equivalent to a pair of directionally unbiased three-ports arranged as shown
on the right.}
\label{diamfig}
\end{figure}

It was shown in Ref. \cite{threeport} that the unbiased multiport described
by Eq. \ref{U3port} provides a physical realization of the abstract
three-point scattering vertex used in several studies of quantum walks on
graphs \cite{fh1,fh2,fh3}. We now take advantage of that equivalence in order
to apply some of those graph-based results to a physically implementable
optical system. In particular, the basic building blocks of the system will
be the units shown on the left in Fig. \ref{diamfig}, whose properties were
studied in \cite{fh1,fh2,fh3}. Each such diamond graph consists of a pair of
three-point scattering vertices connected at two edges, with an additional
phase shift on one connecting edge. The remaining two edges provide
input/output lines. Given the equivalence between the scattering centers and
the unbiased three-ports, this system can be physically implemented by a pair
of unbiased three-ports, as shown on the right in Fig. \ref{diamfig}, with
each graph edge corresponding to an allowed optical path. It is assumed here
that the multiports are very small (effectively pointlike) compared to the
distance $d$ between them.

To simulate SSH-like behavior, each $a$ and $b$ subsite in Fig. \ref{sshfig}
is formed from one such diamond graph, so each unit cell contains four
multiports and two phase shifters. The phase shifts $\phi_a$ and $\phi_b$ in
the two diamonds may be different from each other; they are adjustable
parameters that can be varied independently.

\begin{figure}
\centering
\includegraphics[totalheight=.55in]{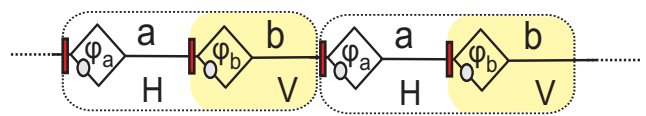}
\caption{Simulating the SSH Hamiltonian with diamond graphs. Each graph is made from two directionally unbiased three-ports, as shown in Fig. \ref{diamfig},
so that each cell (indicated by the dashed curves) is made from four three-ports.
The two diamond graphs at each site may have different internal phase shifts, $\phi_a$ and $\phi_b$.
The red rectangles are phase plates that rotate the polarization by $90^\circ,$ so photons in the shaded areas have vertical polarization (state b),
while the unshaded regions have horizontal polarization (state a).  The size of the
diamond graphs is exaggerated for clarity: they should be small compared to the distance separating them.}
\label{sshv4fig}
\end{figure}

Drawing the simulation system in the form of the abstract diamond graphs, it
then looks as shown in Fig. \ref{sshv4fig}. The red rectangles are phase
plates, rotating the polarization by $90^\circ$. The time unit $T$ is taken
to be the time to go from one diamond graph to the next, or equivalently from
one phase plate to another. Having a photon present in the areas labeled $a$
and $b$, bounded by the phase plates, represent the two substates at each
lattice site. If $\phi_a\ne\phi_b$, the two triangle graphs inside each cell
will have different transmittances. Let $T_a=|t_a|^2$ and $T_b=|t_b|^2$ then
be the transmission probabilities corresponding to the two graphs. In order
to make it easy to measure which subsite the photon is in within a cell, the
phase plates will flip the polarization each time the substate changes. Here,
the polarization simply serves as a convenient bookkeeping device to make
experimental distinguishability of the $a$ and $b$ states easier, and is not
essential to the theoretical development. Photons can be easily coupled in
and out of the system by means of optical switches and optical circulators as
described in more detail in \cite{simham}.

There is one photon collision with a diamond graph per unit time. At each
encounter with one of these graphs, there are amplitudes to either reflect
back from it into the original subsite, or to be transmitted through to an
adjacent subcell. The hopping amplitudes $v$ and $w$ are then given by the
transmission amplitudes of the diamond graphs. Without loss of generality, an
appropriate redefinition of states may always be used to make the amplitudes
real, in which case
\begin{eqnarray}\langle m,a|m,b\rangle \; =\; |t_a | \label{vta}\\
\langle m+1,a|m,b\rangle \; =\; \langle m,b|m+1,a\rangle \; =\; |t_b |
.\label{wtb}
\end{eqnarray}

Transitions in which the photon reflects off the diamond graph and back into
the same subcell give diagonal contributions to the Hamiltonian that simply
shift all of the energies up or down by the same amount; i.e. they define the
zero level of the energy. These terms can therefore be ignored for current
purposes. The remaining terms are those that take a photon from one subcell
to an adjacent subcell in a single time step; in other words, terms of the
form that appear in the Hamiltonian of \ref{sshhamiltonian}. The hopping
amplitudes $v$ and $w$ between subcells are given by the transmission
coefficients of the diamond graphs. Therefore, the Hamiltonian of interest is
\begin{eqnarray}\hat H &=& |t_a | \sum_{m=1}^N \left( |m,b\rangle \langle m,a| +  |m,a\rangle \langle m,b|  \right)\label{sshhamiltonianv2} \\
& & + |t_b |\sum_{m=1}^{N-1} \left( |m+1,a\rangle \langle m,b| +  |m,b\rangle \langle m+1,a|  \right) ,\nonumber\end{eqnarray} where the diamond graph
transmission amplitudes for phase shifts are \cite{fh1,fh2,fh3}:
\begin{equation} t_j(k) =  {{4(1+e^{-i\phi_j})(1-e^{-i(\phi_j+4k)})}\over
{e^{-4ik}(1+e^{-i\phi_j})^2-(3e^{-i(\phi_j+4k)}-1)^2}} ,\label{Teq}
\end{equation} where $j=a,b$.

%

\subsection{Quasi-energies and transmission amplitudes. }

\begin{figure*}
\begin{center}
\subfigure[]{
\includegraphics[scale=.25]{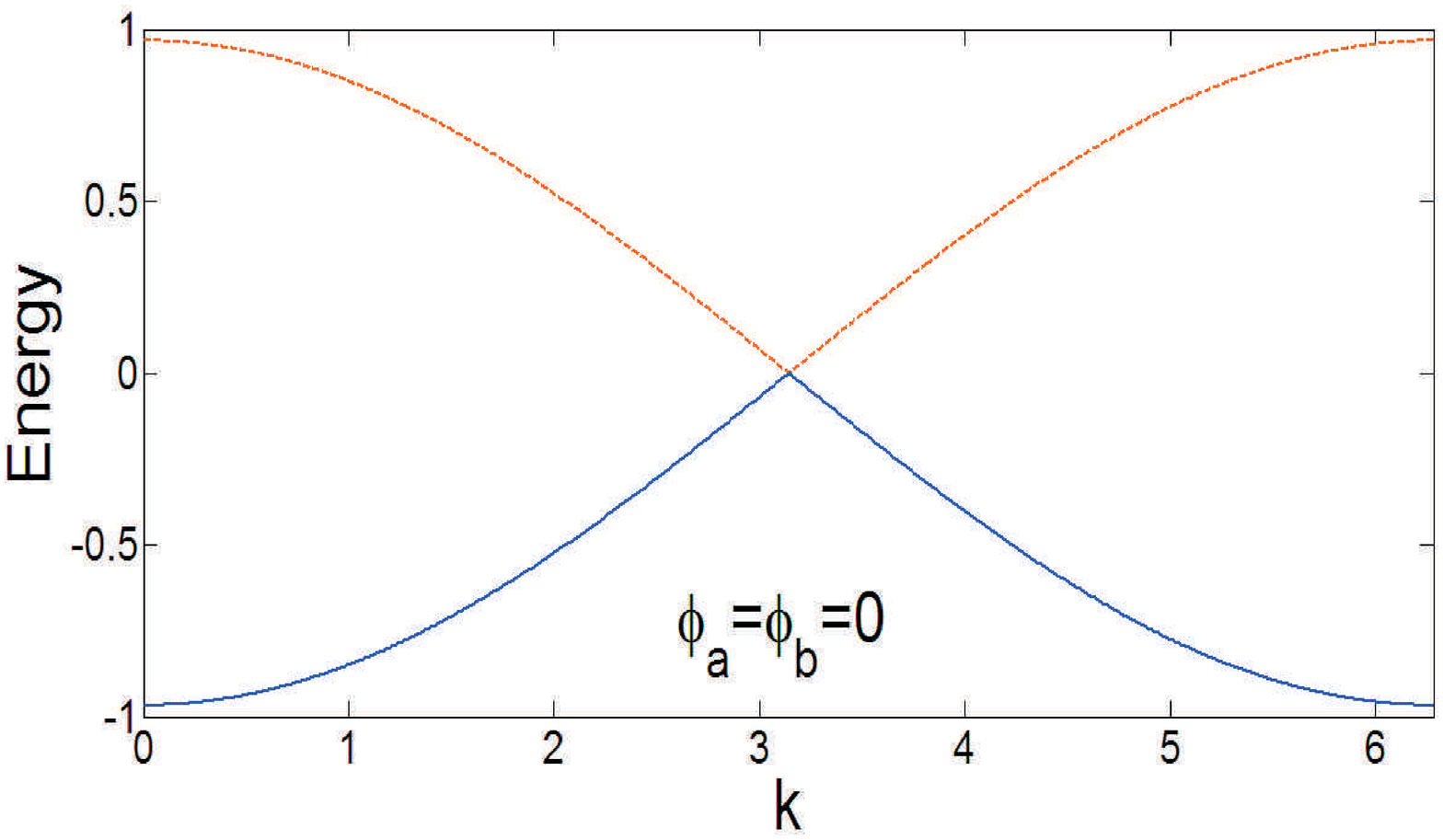}}
 \subfigure[]{
\includegraphics[scale=.25]{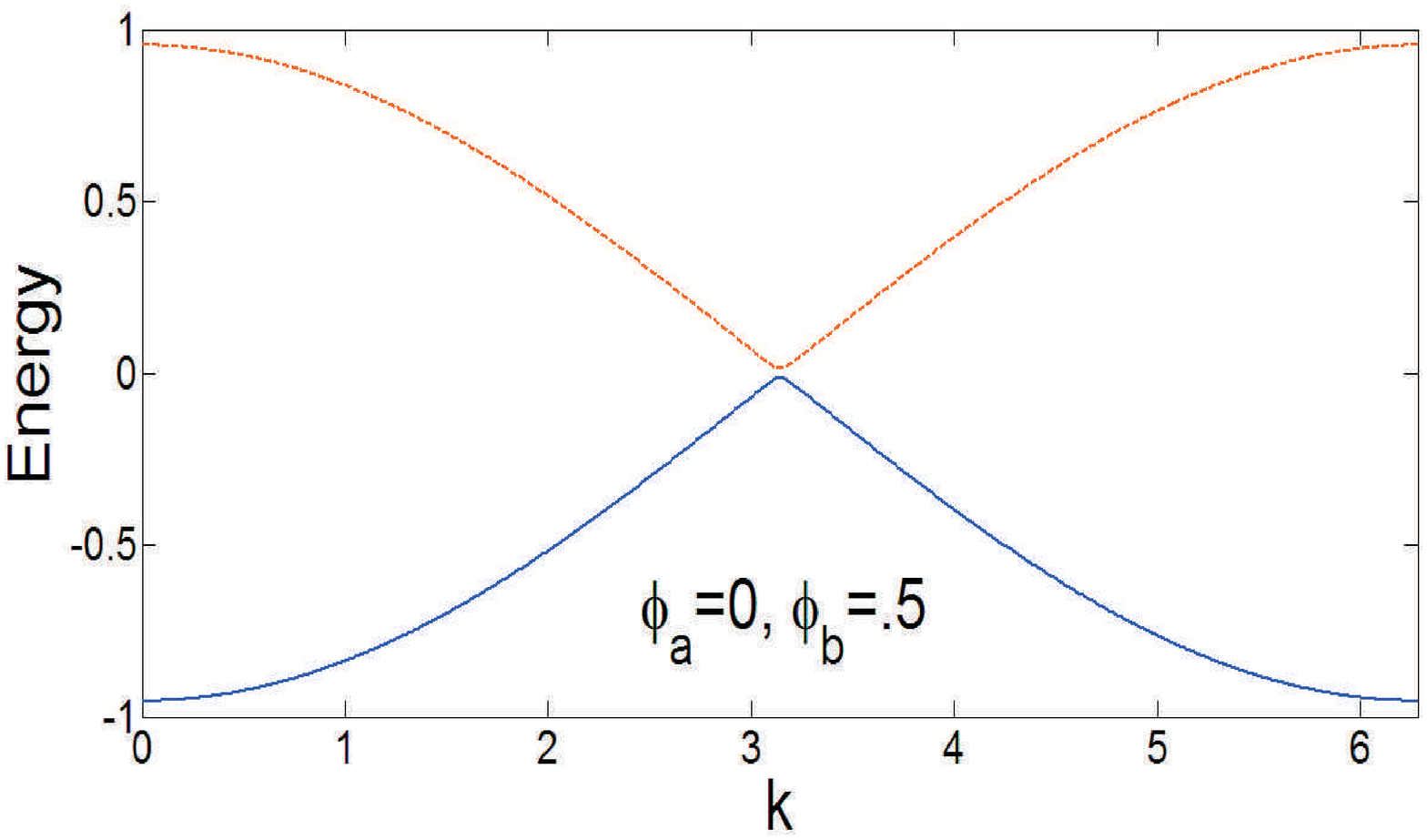}}\\
\subfigure[]{
\includegraphics[scale=.25]{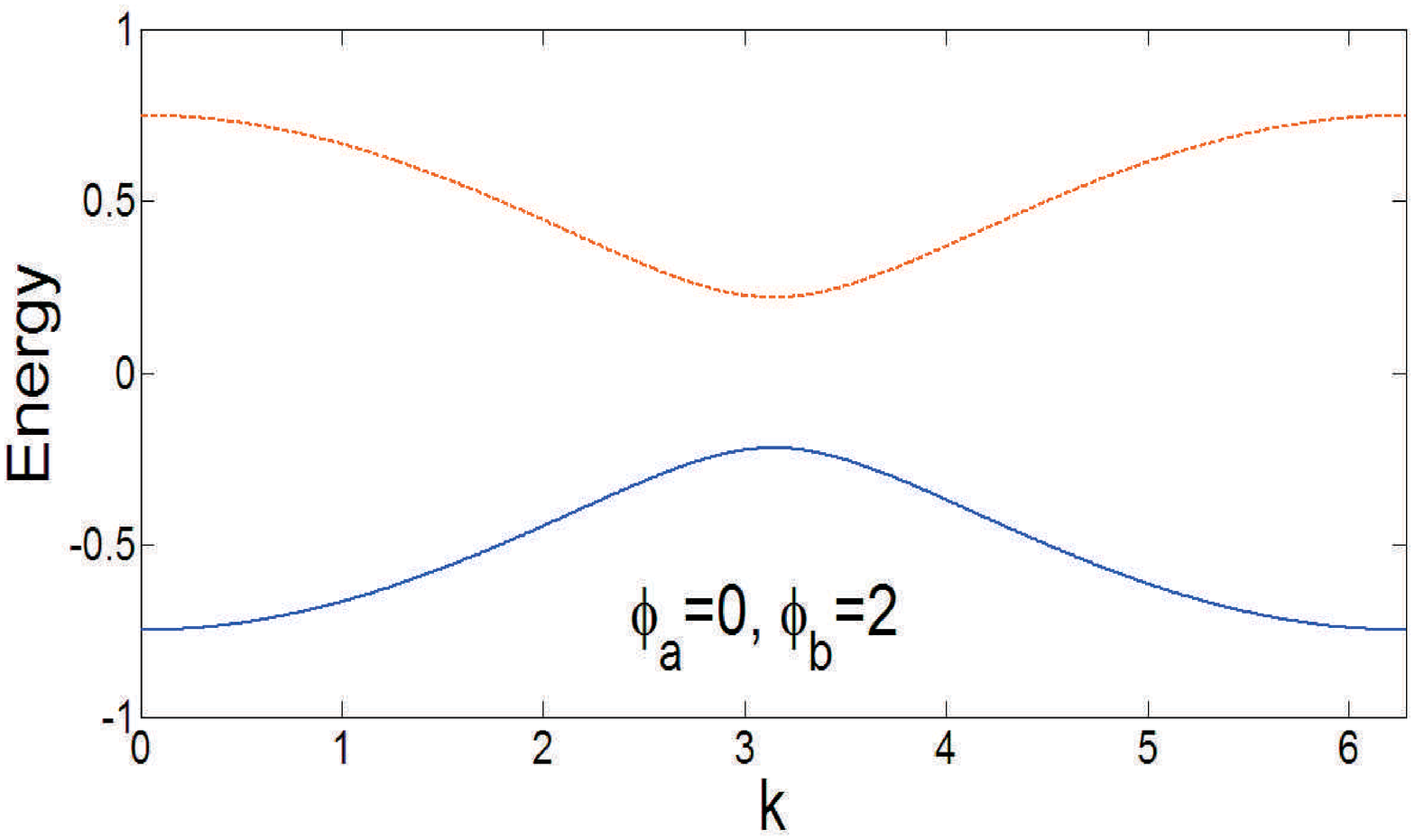}}
 \subfigure[]{
\includegraphics[scale=.25]{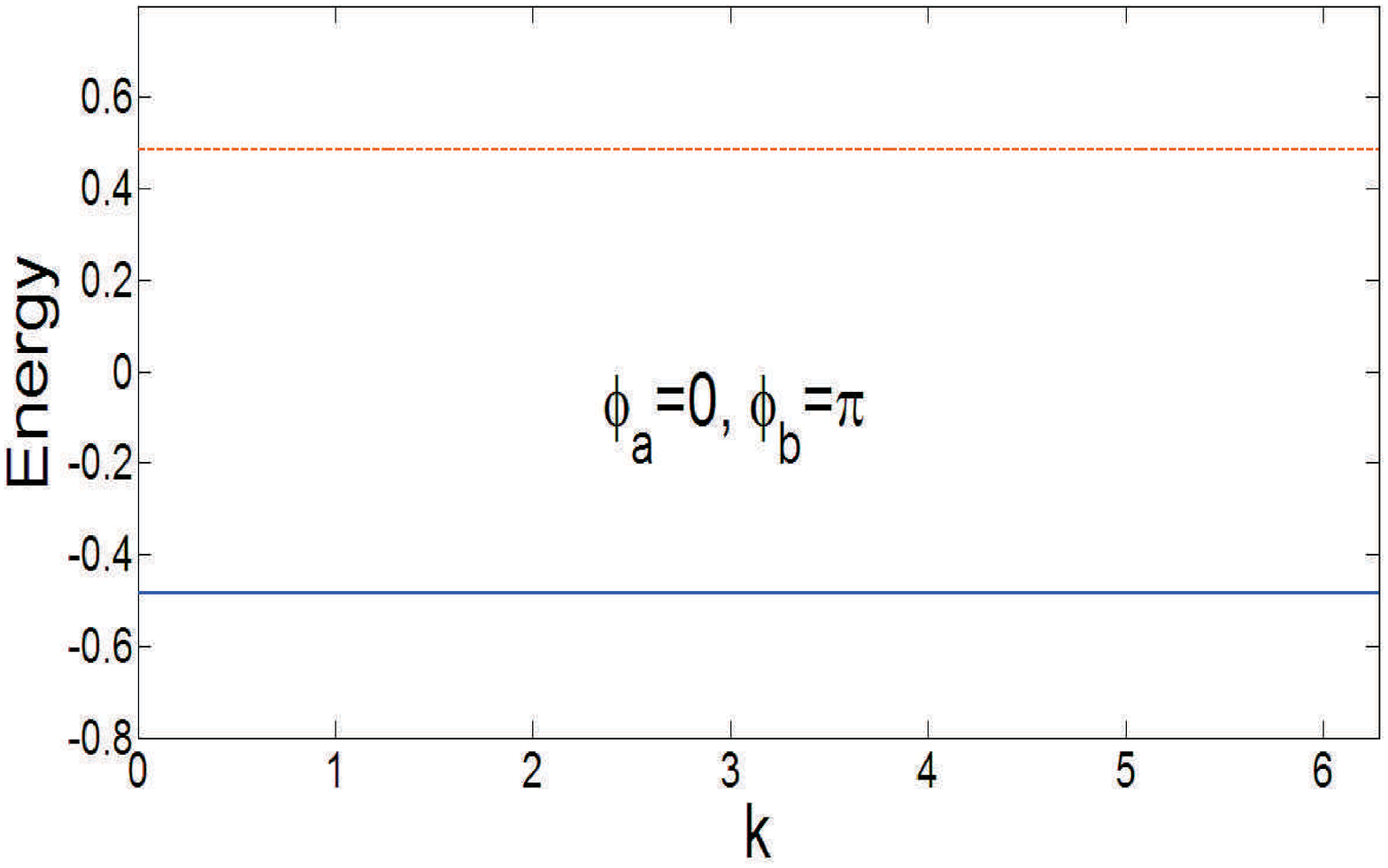}}
\caption{Energy bands for the system shown in Fig. \ref{sshv4fig}.  The
band gap vanishes when the two phases are equal (a) and opens up when the two phase values differ. The gap is still very small in (b), but grows as
$|\phi_a-\phi_b|$ increases (c), reaching its maximum size when $|\phi_a-\phi_b|=\pi$ (d).}\label{Efig}
\end{center}
\end{figure*}

It is convenient to work in quasi-momentum space. Carrying out the Fourier
transform,
\begin{equation}|k\rangle ={1\over \sqrt{N}}\sum_{m=1}^N
e^{imk}|m\rangle,\end{equation}  the momentum-space Hamiltonian is then a
matrix in the $a$-$b$ internal space:
\begin{equation}\hat H =
{1\over N}\sum_k \hat H(k) \; |k\rangle\langle k|
,\end{equation} where
\begin{equation}\hat H(k) = \langle k|\hat
H|k\rangle ={1\over N}\left( \begin{array}{cc}0 & |t_a|+|t_b|e^{-ik}\\
|t_a|+|t_b|e^{ik} & 0\end{array} \right). \end{equation} For each value of
$k$, this has two eigenvalues \begin{equation}E_\pm(k)  =\pm
\sqrt{|t_a|^2+|t_b|^2+2|t_at_b|\cos k} .\end{equation} Plotting $E$ versus
$k$ gives the analog of a band-gap diagram with minimum band gap $\Delta$.
When $v$ and $w$ are independent of $k$, as in the usual SSH case, then
$\Delta =2|v-w|$, but notice that in the present case $v$ and $w$ depend on
momentum via the $k$-dependent transmittances.   In this sense, this is not
the true SSH model, but a slight variant of it, which we might call the
modified SSH (MSSH) model; this is a continuous deformation of the usual SSH
model and so should have topologically identical behavior, as will be
verified below.

The nonvanishing coefficients of the Pauli matrices in Eq. \ref{Handd} are
now
\begin{eqnarray}d_x (k)&=& |t_a(k)|+|t_b(k)|\cos k \label{dx} \\ d_y(k) &=& |t_b(k)|\sin k .\label{dy}\end{eqnarray}
As $k$ goes from $0$ to $2\pi$, $\bm d$ traces out paths labeled by their
winding numbers $\nu$ about the origin. These winding numbers will be
functions of the hopping amplitudes: $\nu (v,w)=\nu (|t_a|,|t_b|)$. Since
$t_a$ and $t_b$ vary only weakly with $k$, it is clear that the loop traced
out by $\bm d(k)$ encloses the origin and has nonzero winding number if
$|t_b|>|t_a|$.

\begin{figure*}
\centering
\includegraphics[totalheight=4.2in]{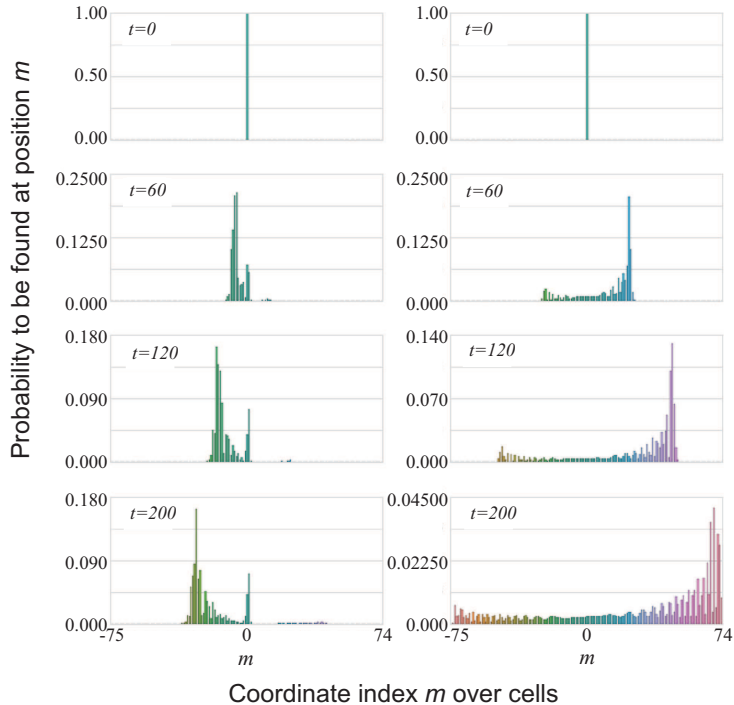}
\caption{Comparison of MSSH quantum walk with topologically protected edge state (left) and MSSH quantum walk with normal ballistic spreading (right) for $0\leq
t \leq 200$. The two walks are simulated for the arrangement shown in Fig. \ref{sshv4fig}. The topology on the left uses $(\phi_{a}=1.5, \phi_{b}=2.5)$ when
$m\leq 0$, giving a winding number $\nu\ne 0$; this joins up with a $\nu=0$ region having phase shifts of $(\phi_{a}=3\pi/4, \phi_{b}=0)$ when $m\geq 0$. The
result is a localized state confined to the boundary of these topologies, indicated by the peak that arises at $m=0$. In contrast, the topology on the right uses
$\phi_{a}=0, \phi_{b}=0$ for all $m$, leading to a ballistic quantum walk in one dimension.} \label{topo}
\end{figure*}



\subsection{Verifying topological behavior: localized boundary states.}
Properties of the system of Fig. \ref{sshv4fig} can be numerically simulated,
and the results verify that the model constructed here has behavior similar
to that expected from the SSH model. Fig. \ref{Efig} shows plots of the
energy levels for different values of the phase shifts $\phi_a$ and $\phi_b$
of the two graphs. When the two phase shifts are equal, the band gap
vanishes. As they begin to differ, a gap opens up and becomes larger with
increasing $|\phi_a-\phi_b|$, reaching a maximum at $|\phi_a-\phi_b|=\pi$, as
Fig. \ref{Efig} shows. The exact shapes of the curves are slightly different
than the pure SSH model (in particular, the value of $k$ that minimizes the
gap clearly shifts horizontally as the parameter changes), but the
qualitative behavior is identical.

Similarly, it is easy to show that different values of the phase shifts allow
solutions with both zero and nonzero winding numbers to occur.  Evaluation of
Eqs. \ref{Teq},  \ref{dx}, and \ref{dy} for a range of $\phi_a$ and $\phi_b$
values readily shows that varying these phases causes the path traced out by
$\bm d$ to shift horizontally and change radius, leading to transitions between
winding numbers $0$ and $1$.
This indicates that different phase values in the two diamond graphs lead to
different topological phases, distinguished by their winding numbers. By
attaching two chains of these graphs with different winding numbers on each
chain there should then arise localized, topologically-protected states at
the boundaries between them \cite{kitagawa,asboth,hasan}.

Fig. \ref{topo} supports this analysis by showing specific conditions under
which topologically protected states can be generated using a network of
multiports. The plots compare two numerical simulations of a single photon
quantum walk on the MSSH model described above for $0\leq t \leq 200$ in
units of $T$. In each case an initially right-moving photon is injected at
subsite $a$ of the $m=0$ position coordinate at $t=0$ (top row). (The
mechanism for physically inserting photon states into the chain is described
in detail in Ref. \cite{simham}.) The left hand column shows the time
evolution over a chain with two different topologies attached to each other
at $m=0$. The right hand side shows time evolution over a chain with uniform
topology.

Specifically, the left column of Fig. \ref{topo} uses subsite phase shifts of
$\phi_{a}=1.5, \phi_{b}=2.5$ for the portion to the left of the origin
($m\leq 0$), giving a winding number $\nu\ne0$ in that region. To the right
of the origin ($m\geq0$), phase shifts of $\phi_{a}=3\pi/4, \phi_{b}=0$ are
used, giving $\nu=0$. The result is a persistent probability of finding the
photon at the boundary between the two topologies, the signature of a
topologically protected edge state \cite{kitagawa}.

For comparison, the right column of Fig. \ref{topo} shows a quantum walk over the MSSH model using no change in phase shifts: $\phi_{a}=0, \phi_{b}=0$ for all
$m$ (positive and negative). As expected, in this case evolution reduces to a standard quantum walk in one dimension, exhibiting ballistic spreading of
probability over the coordinates.


Note that the right side of Fig. \ref{topo} is asymmetric. This is because
the value $\phi_a=\phi_b$ that was used reduces the lattice to a chain of
three-ports with transition matrix of form  Eq. \ref{U3port}; this matrix has
much smaller amplitude to reflect out the input port than to transmit out the
other two, resulting in a strong bias of the photon to continue moving in its
initial direction. For other values of $\phi_a$ or $\phi_b$ (or for other
values of the internal vertex phase $\theta$ of Eq. \ref{general}) this bias
changes or disappears.

It may also be pointed out that the diamond graphs have four bound states \cite{fh2}  when $\phi_a-\phi_b =0$, and none for $\phi_a-\phi_b\ne 0$. These bound states occur in all the diamond graphs at exactly at the parameter values $\phi_a=\phi_b=0$ at which the energy gap closes. This allows the
controlled storage of photons: photons can be stored in the graph or released as the value of $\phi $ is changed.

\section{Conclusions}\label{conclude}

Systems with distinct topological phases are of increasing importance in
condensed matter physics and in quantum computing. The ability to simulate
their properties in a simple manner is therefore of current interest, and the
ability to supply such simulations efficiently using only linear optical
quantum systems would be a useful advance. Here, a method for simulation of
systems in the same topological class as the SSH Hamiltonian was proposed
that for large numbers of time steps requires substantially fewer resources
than the method of \cite{kit3}.

In this paper the focus was on the use of only one-dimensional chains of directionally-unbiased optical three-ports. However, this only scratches the surface of
the possibilities to be examined, since the current approach can be generalized to two- and three-dimensional arrays of $n$-ports with $n>3$. The phase shifts at each vertex of the multiport can also be varied, altering the multiport properties. Thus a rich array
of more sophisticated simulation types remains as-yet unexplored, using the same general methods. These hold promise to provide quantum simulation of a diverse
range of further phenomena.

\section*{Acknowledgements}This research was supported by the National Science Foundation EFRI-ACQUIRE Grant No. ECCS-1640968, NSF Grant No. ECCS- 1309209, and by the Northrop
Grumman NG Next.


\end{document}